# Field-free current-induced magnetization switching of a room temperature van der Waals magnet for neuromorphic computing

Chenxi Zhou, Zhe Guo*, Qifeng Li, Gaojie Zhang, Hao Wu, Jinsen Chen, Rongxin Li, Shuai Zhang, Cuimei Cao, Rui Xiong, Haixin Chang and Long You*

**Abstract**

Spin orbit torque (SOT) has become a promising approach to efficiently manipulate the magnetization switching in spintronic devices. As a main factor to impact the device performance, the high quality interface is essentially desired, which can be readily acquired by using the two-dimensional (2D) van der Waals (vdW) materials. Recently, a 2D ferromagnetic material $Fe_3GaTe_2$ has been discovered to possess the above-room-temperature Curie temperature and strong perpendicular magnetic anisotropy (PMA), providing an excellent candidate to build spintronic devices. On the other hand, an external magnetic field is necessary for the SOT-driven deterministic switching of perpendicular magnetization, which has become a block for the real applications. Here, we realize the field-free SOT switching of $Fe_3GaTe_2$ at room temperature based on the $Fe_3GaTe_2$/MnPt heterostructure. In addition, inspired by the superiority of 2D materials in 3D heterogeneous integration, we explore the potential of our device in the computing in memory (CIM). With the application of the current pulses, the gradual switching of our device at zero field imitates the function of artificial synapse in the convolutional neural network (CNN), achieving a high accuracy (~ 92.8%) pattern recognition. Our work proposes a feasible solution for field-free SOT switching in 2D vdW spintronic devices, which paves the way for applications in magnetic memory and neuromorphic computing.

# Introduction

Magnetic random access memory (MRAM) has attracted tremendous attentions in the emerging computing applications due to its superior characteristics such as nonvolatility, fast speed, high endurance and energy efficiency [1-3]. Compared with spin transfer torque (STT), the spin orbit torque (SOT) switching approach enables the separation of writing and reading path, leading to faster writing speed and higher endurance [4-6]. The switching mechanism of SOT has been regarded as originating from the spin hall effect [7] or Rashba effect [8], which are both strictly related to the interface. In addition, as the core of MRAM, the CoFeB/MgO based magnetic tunnel junction prefers to possess the strong perpendicular magnetic anisotropy (PMA) for miniaturization, which also depends on the interfacial effect [9, 10]. Thus, the excellent interface is desired for constructing high performance spintronic devices. The two-dimensional (2D) magnetic materials intrinsically possess clean surface without dangling bonds, bringing new opportunities for spintronic devices [11-13]. To date, a variety of 2D magnetic materials, such as $CrI_3$ [14], $Cr_2Ge_2Te_6$ [15] and $Fe_3GeTe_2$ [16] have been discovered and widely investigated. However, these materials have low Curie temperature, hindering their actual applications.

Recently, an above-room-temperature 2D ferromagnet $Fe_3GaTe_2$ (FGT) has been discovered [17]. The FGT exhibits high Curie temperature and strong PMA intrinsically. As high as 85% tunneling magnetoresistance (TMR) ratio has been achieved based on FGT/hBN/FGT junction at room temperature [18]. Compared with the interfacial PMA of conventional CoFeB/MgO heterostructure, such intrinsic PMA in FGT facilitates the further miniaturization from the view of thermal stability. A few works about the manipulation of FGT magnetization through SOT have been reported [19-20]. Especially, field-free switching of FGT have been realized by the asymmetric geometry design [21, 22], orbit transport torque [23] and out-of-plane spin polarized current [24-27]. On the other hand, 2D materials offer great potential in 3D heterogeneous integration as different materials can be easily stacked [28]. Meanwhile, they can incorporate a wealth of functionalities for integration such as sensing [29], storage [30-32], computing [33, 34] and energy harvesting [35, 36]. The discovery of 2D magnetic materials brings new opportunities for such integration such as sensing the magnetic fields [37] and potentials in the new computing paradigms [38, 39].

In this work, we focus on the current-induced magnetization switching of 2D FGT and explore its potential in the neuromorphic computing. Firstly, we fabricated FGT/MnPt heterostructure and realized the field-free SOT switching at room temperature. Then the ability to serve as artificial synapse for neuromorphic computing was investigated. The weight modulation behavior was manipulated through applying current pulses to stimulate the gradual switching at zero field. The potentiation/depression behavior was observed under positive/negative pulses. A convolutional neural network (CNN) was constructed based on such artificial synapses and a high recognition accuracy was achieved.

**Results and Discussion**

Figure 1a illustrates the schematic of the FGT vdW atomic structure. The vdW gap is between adjacent Te layers along *c* axis and the FGT is stacked along *c* axis. In each FGT layer the $Fe_3Ga$ heterometallic slab is sandwiched between two Te layers. Our FGT crystal is synthesized by the self-flux method. The FGT nanosheets were mechanically exfoliated with blue membrane tapes and transferred onto a silicon or MgO substrate. Figure 1b shows few layer FGT nanosheet with thickness of 13.5 nm. The typical triangular shape was observed, which are common for 2D materials with hexagonal structures. The Raman spectra (Figure 1c) of FGT nanosheet (21.3 nm thick) presents lattice vibration modes $E_1$ (~94 $cm^{-1}$), $A_1$ (~128 $cm^{-1}$) and $E_2$ (~144 $cm^{-1}$) of Te atoms, which are the features of 2D telluride compounds. To investigate the PMA of FGT nanosheet, the anomalous hall effect (AHE) is employed. We deposited electrodes on top of FGT nanosheet and then fabricated into Hall-bar structure by photolithography and ion milling. The magnetic field was applied perpendicular to the surface of the FGT nanosheets and all the measurements were performed at room temperature. As shown in Figure 1d, the FGT nanosheets presents different magnetic properties with thickness varying. The FGT nanosheet with a thickness of 344 nm displays a loop resembling the one consisting of two phases, which has also been observed in the $Fe_3GeTe_2$ nanosheets [40]. Such a phenomenon may be due to the domain evolution from a soft phase of bulk crystal to a hard magnetic phase. With nanosheet thickness down to 37 and 15 nm, the square loops with remanence ratio of 1 appear. The magnetization flips sharply, similar to a single domain behavior, indicating good PMA.

To study the current-induced magnetization switching of FGT, we first investigate the

FGT/Pt (heavy metal, HM) heterostructure for comparison. The exfoliated FGT nanosheets were transferred onto a silicon substrate with a 300-nm-thick oxide layer, and 6 nm-thick Pt layer were sputtered on the FGT nanosheets to generate spin current (Figure 2a, see fabrication details in Methods). The optical image of the fabricated Hall-bar structure is shown in Figure 2b, and the atomic force microscope (AFM) measurement indicates the thickness of the FGT is around 28 nm. As shown in Figure 2c, the exfoliated FGT has good PMA. The current-induced magnetization switching is conducted through applying current pulses with varying magnitudes along the $x$ axis (Figure 2a), while the Hall resistances ($R_H$) are collected with a smaller reading current after each writing pulse. The SOTs generated from Pt layer switch the perpendicular magnetization of FGT, but with the assistance of an external in-plane magnetic field ($H_x$) to break the symmetry. Figure 2d shows the switching loops with $H_x$ varying from 800 to −800 Oe. The polarity of magnetization switching changes when the direction of $H_x$ reverses, which is the typical attribute of SOT switching. We note that the two magnetization states during the switching are not fully saturated (the switching ratio is around 9.5%), which may be attributed to the effects of Joule heating [19]. However, without the help of the external $H_x$, no magnetic switching can be observed in such structure.

Next we investigate the current-induced magnetization switching in the FGT/MnPt heterostructure in which the MnPt serves as the SOT source and meanwhile contributes to the field-free switching. The fabrication process is similar to the previous one for the FGT/Pt heterostructure. After transferring the FGT nanosheets onto the MgO substrate, we deposited MnPt film on top of the nanosheets by sputtering at room temperature (Figure 3a). We examined the X-ray diffraction (XRD) spectra of the pure MnPt film which was directly deposited on the MgO substrate with the same deposition conditions. The representative $\theta$−$2\theta$ pattern (Figure S1) shows (002) peak assigned to the MnPt layer, indicating its preferred orientation growth. The X-ray reflectivity (XRR) spectrum (Figure S1) shows multiple oscillations which means the smooth interface. However, the absence of (001) peak of MnPt layer implies that the deposited MnPt film has a disordered crystal structure. Figure 3b shows the optical image of the fabricated device. The nanosheet thickness is 18 nm and the corresponding AHE loop also shows good PMA (Figure 3c). The AHE response of the pure MnPt film (Figure S2) shows negligible variations, which is consistent with its collinear antiferromagnetic structure. The current-induced switching loops with the varying $H_x$ are exhibited in Figure 3e. The loops in the left (right) panel were collected with $H_x$ varying from +500 Oe (-500 Oe) to -500 Oe (+500 Oe). Obviously, different from the FGT/Pt heterostructure, the deterministic switching in the absence of an external field has been demonstrated in FGT/MnPt bilayer. Moreover, the switching polarity was not influenced by the history of the applied $H_x$.

To validate the reliability of the field-free switching characteristic in the FGT/MnPt heterostructure, we have fabricated plenty of samples with the same structure, all of which exhibit the field-free SOT switching behavior (Figure S3). In contrast, no SOT induced magnetization switching was observed in the absence of an assistant magnetic field in the other fabricated FGT/Pt system. These results prompt us to discuss the potential mechanism that may contribute to the field-free SOT induced magnetization switching observed in FGT/MnPt system, such as stray field generated from the MnPt film and the interface effect between MnPt and FGT layer. We first measure the *M-H* loop of the MnPt film by employing a physical property measurement system (PPMS, Quantum Design), which exhibits a negligible remanent magnetization (Figure 3d), indicating that the SOT induced magnetization switching due to stray filed is negligible. Possibly, the exchange coupling between the Mn magnetic moments at the bottom MnPt surface and the FGT layer might cause the tilting of the magnetization of FGT along the in-plane direction, enabling the field-free SOT induced magnetization [41]. In addition, the out-of-plane spin polarization, which could be originated from the out-of-plane tilting of the Mn magnetic moments induced by the enhancement of the extrinsic spin Hall effect in Mn due to HM Pt doping, may also lead to the field-free switching [42]. Based on the analysis above, the main candidate mechanisms can be listed as follows: i) the out-of-plane polarized spin current generated by the exchange coupling effect at the FGT/MnPt interface, ii) the out-of-plane tilting of the Mn magnetic moments at the MnPt surface. However, more detailed theoretical and experimental studies are requisite to further clarify these results.

Then we explore the potential of FGT in the application of neuromorphic computing. As a fundamental unit of neuromorphic computing, the artificial synapse is responsible for the connectivity between pre- and post-neuron. The weight representing the connection strength prefers to be modulated continuously and linearly in a nonvolatile manner, which is also called the memristive behavior. In actual applications, the conductance of a memristive device is extracted as the synaptic weight [43]. In this work we exploit the current pulses to stimulate the device (Figure 4a) and update the conductance through changing the pulse numbers, which is the most popular manner for weight updating. As shown in Figure 4b, 120 positive (negative) current pulses with a fixed amplitude (6 mA) and pulse width (100 μs) are applied to stimulate the device at zero field. As a result, the normalized $R_H$ collected from the device demonstrates that the positive (negative) pulses cause the decrease (increase) of $R_H$, corresponding to the depression (potentiation) of the synaptic weights. The gradual domain nucleation or domain wall motion could account for the changes of $R_H$ [39, 42]. The deviation from the ideal linearity should be associated with nonuniform distribution of domain switching energy barrier in the crossbar area, which may be due to the random distributed pinning sites.

Subsequently, we apply the measured memristive data in the training process of the CNN. To date, CNN plays an important role in computer vision field, including image processing, pattern recognition and object detection. CNN consists of multiple convolutional layers, pooling layers and fully-connected layers. Convolutional layer is the core part of CNN, which uses several convolution units to extract different features of the input. Pooling layer is utilized to reduce the dimension of the feature map, while the fully-connected layer combines and classifies the features to make the final prediction. In this work, we trained a CNN based on the data extracted from our FGT/MnPt device (Figure 4c) to recognize a handwritten dataset (four characters, "h", "u", "s" and "t"). The CNN we used has 2 convolutional layers, 2 max pooling layers and 1 fully-connected layer. The images are 28 × 28 pixels in size with 8,000 images in the training set and 4,000 images in the testing set. In the first convolutional layer, 4 convolution kernels with size of 3 × 3 are used while the second convolutional layer has 8 convolution kernels. The max pooling size is 2 × 2. There are 12 neurons in the fully-connected layer, and the ReLU function is performed as the activation function. The Softmax function is utilized in the output layer to generate a four-class classification outcome for the dataset, where the predicted label category corresponds to the output neuron with the highest probability. The batch size of each training epoch is set to 32. Figure 4d shows that the recognition accuracy can reach up to 92.8%.

**Conclusion**

In summary, we have demonstrated robust field-free SOT induced magnetization switching in the FGT/MnPt heterostructure at room temperature. In addition, our device presents the memristive switching behavior by applying current pulses to continuously modulate the magnetization states. A CNN based on our devices are constructed, achieving a high recognition accuracy up to 92.8%. We believe that our work paves the way to the further study of vdW-based spintronic devices and their application in memory and neuromorphic computing.

**Methods**

**Device Fabrication.** We obtained few-layered FGT flakes with a freshly cleaved surface on MgO substrate by tape exfoliation method. Then, the sample was qucikly loaded into the sputtering machine within 5 minutes and a 10-nm-thick MnPt (or a 6-nm-thick Pt) layer was deposited on the FGT surface, enabling the obtained FGT/MnPt (or FGT/Pt) bilayer was air-stable and high-qualified. The FGT/MnPt (or FGT/Pt) bilayer was patterned into hallbar structure by laser direct writing, and the areas exposed to the air were etched by Ar ion-beam etching.

**Characterization.** All the transport measurements were performed through a home-made magnetoelectric transport characteristics testing system. The electronmagnet (Eastchanging Inc.) can provide a stable magnetic field up to 2 T. Writing and reading current was applied by using a Keithley 6221 current source and the hall voltage across the device was collected using a Keithley 2182A nanovoltmeter. Anomalous hall effect measurements with field sweeps were performed using a testing current of 100 μA. For the SOT-induced switching measurements, the writing current with a 100-ms duration time was followed by 10-times reading pulses (make the average to reduce errors). The measurement of the *M-H* loops for pure MnPt film was performed using PPMS system.

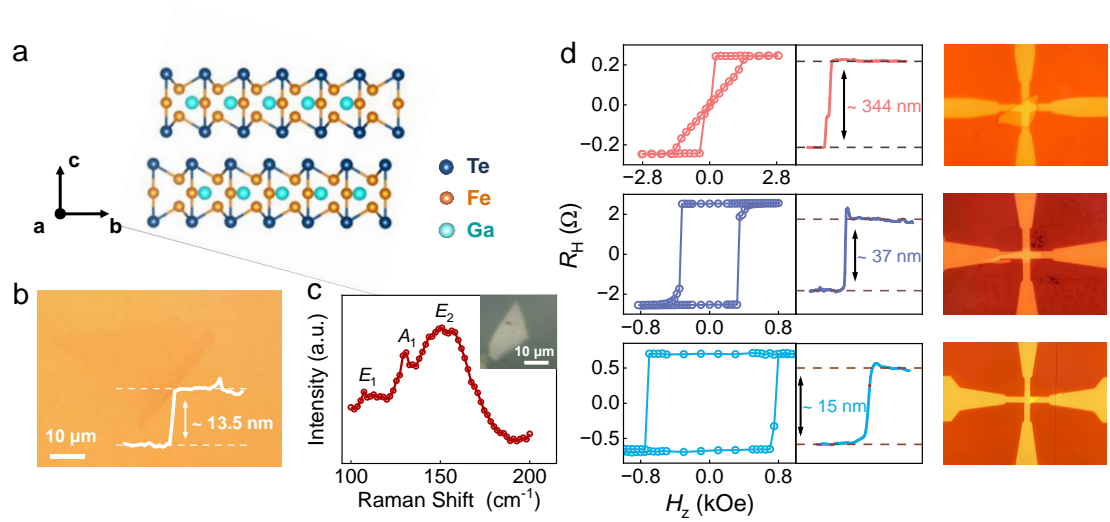

**Figure 1.** Schematic view and characterizations of 2D FGT nanoflakes. (a) Schematic view of the atomic structure of FGT. (b) Optical image of a mechanically exfoliated FGT nanoflake. Scale Bar: 10 μm. AFM topography shows the thickness of 13.5 nm. (c) Raman spectrum of a FGT nanoflake using 532 nm light source. (d) AHE hysteresis loops for FGT nanoflakes with various thicknesses at room temperature and the corresponding optical images of the devcies.

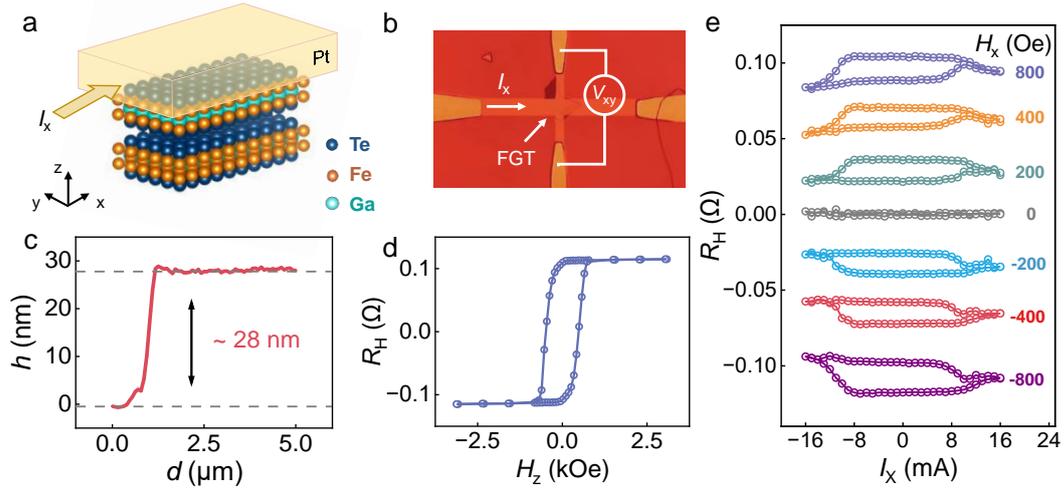

**Figure 2.** SOT induced magnetization switching in the FGT/Pt heterostructure at room temperature. (a) Schematic diagram of the FGT/Pt bilayer device. The yellow arrow represents the current flowing in the Pt layer along the *x* axis, which generates a spin current in the *z* direction. The spin current is injected into the FGT layer and changes its magnetization state. The thickness of the Pt layer is 6 nm. (b) Optical image of the FGT/Pt device. (c) AFM testing result for the FGT/Pt device corresponding to (b), which shows a thickness of 28 nm. (d) AHE loop measured from the fabricated FGT/Pt device at room temperature. (e) Current induced perpendicular magnetization switching under different in-plane magnetic fields. No switching has been observed under zero external magnetic field.

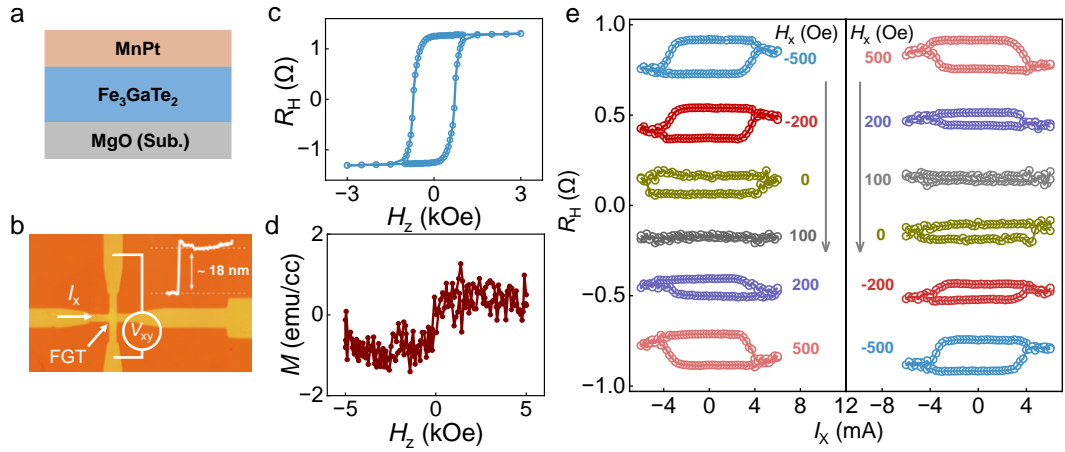

**Figure 3.** Field-free SOT driven magnetization switching in FGT/MnPt heterostructure at room temperature. (a) Schematic diagram of the FGT/MnPt heterostructure. Spin current generated from MnPt layer is injected into FGT and changes its magnetization states. The thickness of the MnPt layer is 10 nm. (b) The optical image, AFM measurement and (c) AHE loop of the FGT/MnPt device. (d) *M-H* loop of the pure MnPt film. (e) Current induced magnetization switching of FGT/MnPt device under different in-plane magnetic fields which range from -500 Oe to 500 Oe and 500 Oe to -500 Oe, respectively. Field-free magnetization switching is obviously observed.

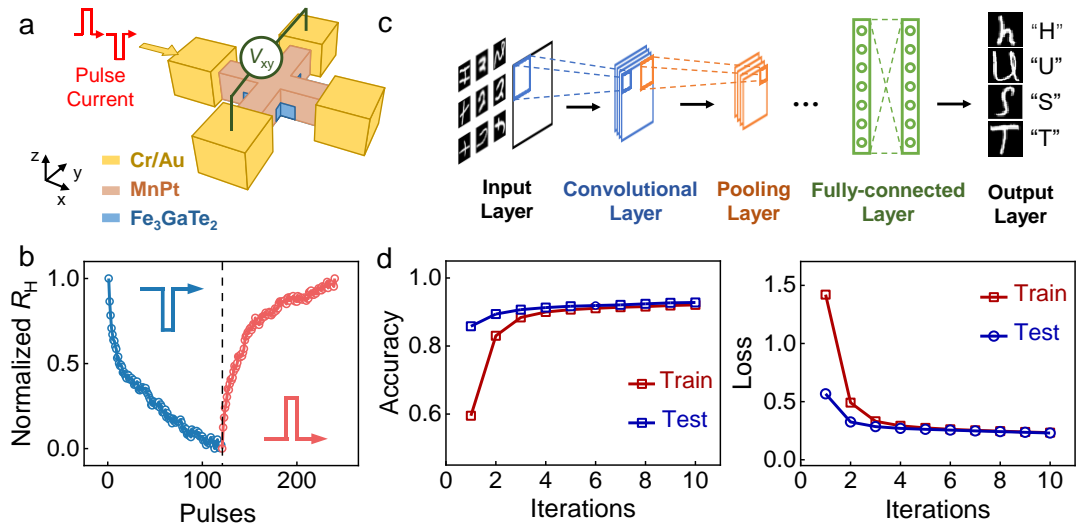

**Figure 4.** Memristive switching of FGT/MnPt device and its application in CNN. (a) Measurement setup of the memristive test. At zero field, positive or negative current pulses are applied along $x$ axis and $V_{xy}$ is collected. (b) Pulse number dependence of $R_H$. Positive (+6 mA) and negative (−6 mA) pulse trains are alternately applied. Each pulse has a duration of 100 μs. $R_H$ has been normalized to [0,1] (c) Schematic of CNN. The trained CNN can recognize four handwritten letters, "h", "u", "s" and "t". (d) Epoch accuracy curve and loss curve of the CNN based on the FGT/MnPt device.

**Author Information**

**Corresponding Authors**

Long You – School of Integrated Circuits & Wuhan National Laboratory for Optoelectronics & Key Laboratory of Information Storage System, Ministry of Education of China, Huazhong University of Science and Technology, Wuhan 430074, China; Shenzhen Huazhong University of Science and Technology Research Institute, Shenzhen 518000, China; Email: lyou@hust.edu.cn.

Zhe Guo – School of Microelectronics, Hubei University, Wuhan 430062, China; Email: guozhe@hubu.edu.cn.

**Authors**

Chenxi Zhou – School of Integrated Circuits, Wuhan 430074, China;

Qifeng Li – Key Laboratory of Artificial Micro- and Nano-structures of Ministry of Education, School of Physics and Technology, Wuhan University, Wuhan 430072, China;

Gaojie Zhang – Center for Joining and Electronic Packaging, State Key Laboratory of Material Processing and Die & Mold Technology, School of Materials Science and Engineering, Huazhong University of Science and Technology, Wuhan 430074, China;

Hao Wu – Center for Joining and Electronic Packaging, State Key Laboratory of Material Processing and Die & Mold Technology, School of Materials Science and Engineering, Huazhong University of Science and Technology, Wuhan 430074, China;

Jinsen Chen – School of Microelectronics, Hubei University, Wuhan 430062, China;

Rongxin Li – School of Integrated Circuits, Wuhan 430074, China;

Shuai Zhang – School of Integrated Circuits, Wuhan 430074, China;

Cuimei Cao – School of Integrated Circuits, Wuhan 430074, China;

Rui Xiong – Key Laboratory of Artificial Micro- and Nano-structures of Ministry of Education, School of Physics and Technology, Wuhan University, Wuhan 430072, China;

Haixin Chang – Center for Joining and Electronic Packaging, State Key Laboratory of Material Processing and Die & Mold Technology, School of Materials Science and Engineering, Huazhong University of Science and Technology, Wuhan 430074, China

**Author Contributions**

C.Z. and Z.G. conceived the idea, L. Y. supervised the project. G.Z., H.W. and H. C. grew samples. C.Z fabricated the device. C.Z. and Z.G. performed electrical transport measurements and analyzed the results together with C.C. and R.L. J.C. performed the simulation. Q.L. and R.X. performed the XRD, XRR and PPMS measurements. C.Z. and S.Z. performed AFM and Raman measurements. C.Z., Z.G. and L. Y. wrote the manuscript with input from all authors. All authors discussed the results and commented on the manuscript.

**Acknowledgment**

This work was supported by the National Natural Science Foundation of China (NSFC Grant Nos. 12327806, 62074063, 52401301, 61821003, 61904060), Shenzhen Science and Technology Program (Grant No. JCYJ20220818103410022), National Key Research and Development Program of China (Grant No. 2020AAA0109005, 2023YFB4502100).